\documentclass{article}

\usepackage{natbib}
\setcitestyle{numbers,square}

\usepackage[preprint]{neurips_2024}

\usepackage[utf8]{inputenc} 
\usepackage[T1]{fontenc}    
\usepackage{hyperref}       
\usepackage{url}            
\usepackage{booktabs}       
\usepackage{amsfonts}       
\usepackage{nicefrac}       
\usepackage{microtype}      
\usepackage{xcolor}         
\usepackage{amsmath}
\usepackage{graphicx}
\usepackage{subcaption}
\usepackage{colortbl}
\usepackage{multirow}
\usepackage{float}

\title{From Chaos to Clarity: 3DGS in the Dark}

\author{%
  Zhihao Li\thanks{Both authors contributed equally to this research.} \quad
  Yufei Wang\footnotemark[1] \quad
  Alex Kot \quad
  Bihan Wen\thanks{Corresponding author. Email: \texttt{bihan.wen@ntu.edu.sg}} \\
  Department of EEE, Nanyang Technology University, Singapore \\
  \textcolor{red}{\url{https://lizhihao6.github.io/Raw3DGS}}
}

\begin{document}

\maketitle

\begin{abstract}
    Novel view synthesis from raw images provides superior high dynamic range (HDR) information compared to reconstructions from low dynamic range RGB images. However, the inherent noise in unprocessed raw images compromises the accuracy of 3D scene representation. Our study reveals that 3D Gaussian Splatting (3DGS) is particularly susceptible to this noise, leading to numerous elongated Gaussian shapes that overfit the noise, thereby significantly degrading reconstruction quality and reducing inference speed, especially in scenarios with limited views. To address these issues, we introduce a novel self-supervised learning framework designed to reconstruct HDR 3DGS from a limited number of noisy raw images. This framework enhances 3DGS by integrating a noise extractor and employing a noise-robust reconstruction loss that leverages a noise distribution prior. Experimental results show that our method outperforms LDR/HDR 3DGS and previous state-of-the-art (SOTA) self-supervised and supervised pre-trained models in both reconstruction quality and inference speed on the RawNeRF dataset across a broad range of training views. Code can be found in \url{https://lizhihao6.github.io/Raw3DGS}.
\end{abstract}

\section{Introduction}

Novel view synthesis (NVS) is fundamental to 3D vision, with extensive applications in virtual and augmented reality (VR/AR)~\cite{li2022rt, li2023instant}, autonomous driving~\cite{li2023read, xu2024regulating}, and 3D asset creation~\cite{metzer2023latent, deng2022gram, lin2023magic3d}. Neural Radiance Fields (NeRF)~\cite{mildenhall2021nerf} have revolutionized this field by rendering colors through the accumulation of RGB values along sampling rays, employing an implicit MultiLayer Perceptron (MLP) representation. Typically, this method uses low dynamic range (LDR) RGB images processed by image signal processing (ISP) modules, leading to a significant loss of crucial scene details, especially in high-contrast areas like highlights and shadows, which can degrade performance in high dynamic range (HDR) environments such as tunnels, sunsets, or dimly lit scenes. Moreover, reliance on ISP-processed RGB images restricts post-capture color and tone adjustments, presenting significant challenges for photographers and modelers during post-production.

In contrast, raw images before ISP offer a higher dynamic range and preserve more scene information. Recent research has indicated that utilizing raw images can significantly enhance the performance of downstream computer vision tasks in complex lighting conditions~\cite{li2024efficient} and offer greater flexibility in post-production adjustments~\cite{zhang2019synthetic,eilertsen2015real}. Building on this advantage, RawNeRF~\cite{rawnerf} first employed raw images as the optimization target in NeRF, achieving marked improvements over traditional RGB-based LDR NeRF approaches. However, RawNeRF's reliance on implicit 3D representation is computationally demanding, requiring up to 48 hours to train a single scene and about one minute to render a single view, which limits its practicality for real-time applications.

\begin{figure}[t]
    \centering
    
    \begin{subfigure}[b]{0.29\textwidth}
        \centering
        \includegraphics[width=\textwidth]{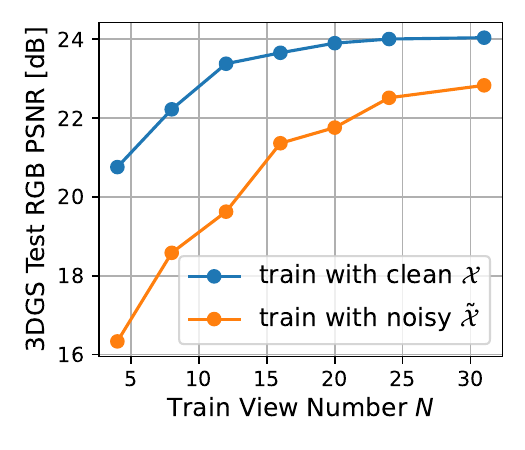}
        \caption{Impact of noise on PSNR}
    \end{subfigure}\label{fig:noise_clean_curve}
    \hfill
    \begin{subfigure}[b]{0.29\textwidth}
        \centering
        \includegraphics[width=\textwidth]{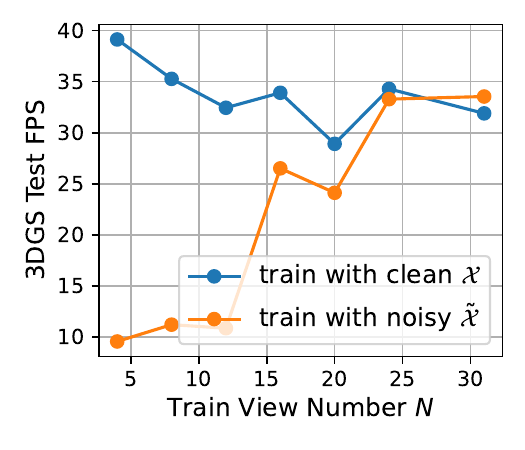}
        \caption{Impact of noise on FPS}\label{fig:fps_curve}
    \end{subfigure}
    \hfill
    \begin{subfigure}[b]{0.37\textwidth}
        \centering
        \includegraphics[width=\textwidth]{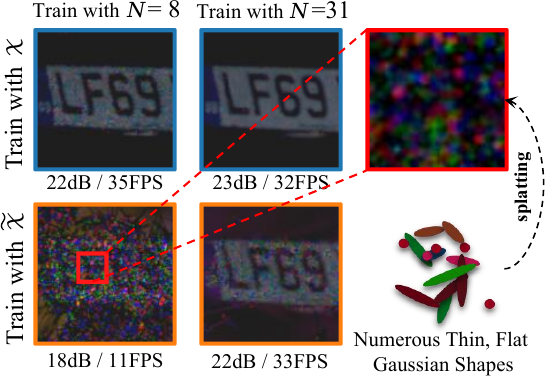}
        \caption{Visualization of test views}\label{fig:test_view_visual}
    \end{subfigure}

    \caption{
        Comparative analysis of 3DGS trained with clean raw images, denoted $\mathcal{X}$, versus noisy raw images, denoted $\tilde{\mathcal{X}}$, across various training view counts $N$. The clean raw images, captured in daylight, are selected from the RawNeRF dataset~\cite{rawnerf}. The noisy raw images are generated from these clean images using the noise model from PMN~\cite{feng2022learnability} with calibrated camera noise parameters. (a) Training with noisy raw images results in decreased PSNR in the test views, with a widening performance gap as the number of training views is reduced. (b) The rendering speed (FPS) shows a similar trend to PSNR. (c) Test view visualizations show that training with noisy images causes 3DGS to produce numerous thin, flat Gaussian shapes, leading to visual artifacts and reduced FPS, especially with fewer training views.
    }\label{fig:cover}
    \vspace{-1em}
\end{figure}

Very recently, the advance on 3D Gaussian Splatting (3DGS)~\cite{kerbl3Dgaussians} offers an explicit 3D representation that employs a set of learnable 3D Gaussian points to depict color, shape, and opacity, thereby enabling real-time novel view rendering. However, utilizing raw images directly as the optimization target in 3DGS can introduce significant artifacts and adversely affect rendering speed, especially in scenarios with limited training views—a common challenge in real-world applications, as depicted in Fig.~\ref{fig:cover}. These challenges arise primarily from the inherent noise in raw images, which is more pronounced than in RGB images due to the absence of noise reduction and smoothing processes typically performed by ISPs. As demonstrated in Fig.~\ref{fig:noise_model_pmn}, the presence of noise in raw images is an inevitable consequence of physical phenomena such as the photoelectric effect and hardware limitations like dark current leakage. Unlike NeRF, where the Multilayer Perceptron (MLP) acts as a low-frequency filter~\cite{zhu2024disorder} to mitigate high-frequency noise, 3DGS tends to produce numerous thin, elongated Gaussians to represent noise, as seen in Fig.~\ref{fig:test_view_visual}.

To address these challenges, we propose a novel self-supervised framework that jointly denoises and constructs HDR 3DGS representations within noisy raw images using a noise robust reconstruction loss. Specifically, our method utilizes a noise extractor to predict the underlying noise in raw images, constrained by a predefined noise model. Simultaneously, the 3DGS branch is optimized towards pseudo noise-free raw images. This dual-branch approach allows the noise extractor to accurately predict and separate noise from the raw images. 
Compared to using the standard RawNeRF loss function with conventional 3DGS, our approach significantly reduces noise artifacts in rendered views and enhances the rendering speed. Our contributions can be summarized as follows:

\begin{itemize}
    \item We discover that noisy raw images significantly degrade the rendering quality and speed of 3DGS, especially in scenarios with limited training views. We provide a detailed analysis of how noise impacts the optimization of 3DGS, and we model its relationship with the number of training views and the noise distribution.
    \item We propose a novel self-supervised framework that incorporates a noise robust reconstruction loss. This framework leverages a physical-based noise model to jointly denoise and enhance the HDR representation of 3DGS within noisy raw images.
    \item Our method substantially outperforms conventional 3DGS methods that employ RawNeRF loss, along with both self-supervised and pre-trained supervised denoisers, in terms of rendering quality and processing speed across various training view counts.
\end{itemize}


\section{Related work}

\subsection{Neural Radiance Fields and 3D Gaussian Splatting}

Recently, Neural Radiance Fields (NeRF)~\cite{mildenhall2021nerf} have spearheaded a significant advancement in novel-view synthesis by reconstructing 3D scenes. NeRF models achieve scene representation by optimizing coordinate-based multi-layer perceptrons (MLPs) to estimate color and density values through differentiable volume rendering. Typically, NeRF utilizes tone-mapped low dynamic range (LDR) images as inputs. These images are processed by image signal processing (ISP), which, while reducing noise and smoothing details, also maps high dynamic range (HDR) to LDR, often clipping highlights and shadows. To counter these limitations, RawNeRF~\cite{rawnerf} modified NeRF with a scaling loss to directly train on linear raw images, thus preserving the scene's full dynamic range. However, like most previous NeRF-based methods, RawNeRF is computationally demanding, requiring up to 48 hours to train a single scene and about one minute to render a single view, which constrains its usability in real-time applications. To accelerate rendering, various methods have been explored, including space discretization~\cite{fridovich2022plenoxels, garbin2021fastnerf, hedman2021baking, reiser2021kilonerf, takikawa2021neural, wu2022scalable, yu2021plenoctrees}, codebooks~\cite{takikawa2022variable}, and hash tables~\cite{muller2022instant}. Yet, these approaches still require a substantial number of queries to render a single pixel, rendering them unsuitable for real-time applications. More recently, 3D Gaussian Splatting (3DGS)~\cite{kerbl3Dgaussians} introduced a novel approach by explicitly representing scenes with learnable 3D Gaussians. Utilizing differentiable splatting and tile-based rasterization, 3DGS enables real-time novel view rendering. However, the initial 3DGS models did not achieve rendering quality comparable to NeRF. To address this, Scaffold-GS~\cite{kerbl3Dgaussians} introduces an anchor-level feature to capture correlations between adjacent Gaussian points, which significantly enhances the rendering quality of 3DGS, achieving comparable results to NeRF while maintaining real-time rendering speeds.

\subsection{Raw image and video denoising methods}

Raw image denoising is critical in both photography~\cite{hasinoff2016burst} and scientific imaging~\cite{levin2020remote, joens2013helium}. Traditional ISPs typically employ self-similarity-based methods such as BM3D~\cite{dabov2007image} to reduce noise. Recent advancements such as SID~\cite{chen2018learning} and SIDD~\cite{abdelhamed2018high} demonstrate that data-driven, deep-learning approaches can surpass these traditional techniques. Innovatively, noise model-based methods like the ELD~\cite{wei2020physics} have extended denoising capabilities into extreme low-light scenarios. It challenges conventional Gaussian noise models by introducing a row-based noise model that uses the Tukey Lambda distribution, which more accurately captures long-tail noise patterns in low-light conditions. The PMN~\cite{feng2022learnability} further investigates Fixed Pattern Noise in dark frames, enhancing alignment with the Poisson-Gaussian (P-G) distribution. These methods, however, often rely on large datasets of paired or clean images for training, as addressed by Neighbor2Neighbor~\cite{huang2021neighbor2neighbor}, which trains denoising models solely on noisy images without clean counterparts. Building on the Neighbor2Neighbor approach, LGBPN~\cite{wang2023lg} introduces more diverse masks to denoise spatially relevant noise. Nevertheless, Neighbor2Neighbor and LGBPN lack integration of a precise noise model or a physical prior of the 3D world, aspects our framework incorporates by merging a noise model with 3DGS to effectively separate noise from noisy images, eliminating the need for clean images. Complementing image-based methods, Yue et al.~\cite{yue2020supervised} expanded the scope to raw video. They introduced a dataset specifically curated for raw video denoising, along with multi-frame-based networks that leverage temporal information across frames to enhance the denoising process.

\section{Preliminaries}

In this section, we describe the noise model that we have adopted to regularize the extracted noise from raw images, provide an overview of the background of 3D Gaussian Splatting (3DGS)\cite{kerbl3Dgaussians}, and review the RawNeRF scaling loss\cite{rawnerf} for handling the wide dynamic range in raw images.

\begin{figure}[t]
    \centering
    
    \begin{subfigure}[b]{0.48\textwidth}
        \centering
        \includegraphics[width=\textwidth]{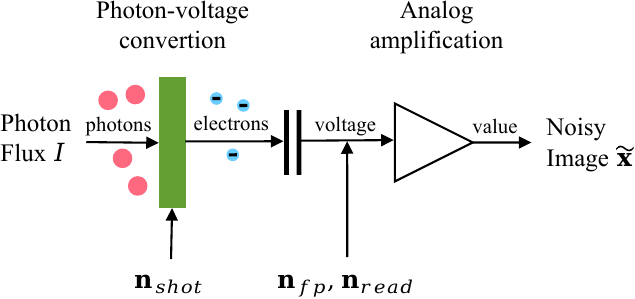}
        \caption{Noise is inevitable in the imaging process}\label{fig:noise_model_pmn}
    \end{subfigure}
    \hfill
    \begin{subfigure}[b]{0.48\textwidth}
        \centering
        \includegraphics[width=\textwidth]{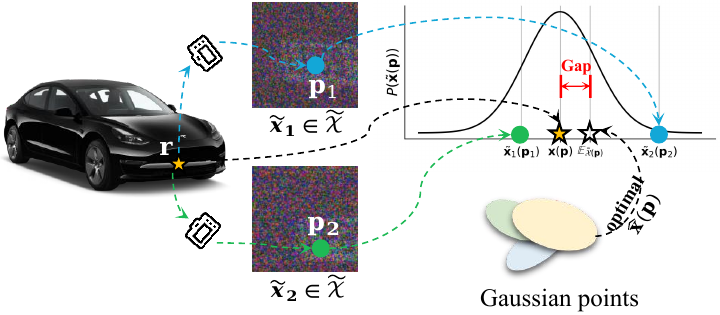}
        \caption{Impact of noise on the 3DGS optimal target}\label{fig:impact_noise}
    \end{subfigure}

    \caption{
        An illustration of how prevalent noise in raw images impacts the 3DGS optimization. (a) The imaging process inherently introduces additive noise at various stages due to physical principles and hardware limitations, represented as $\tilde{\mathbf{x}} = \mathbf{x} + \mathbf{n}$, where $\tilde{\mathbf{x}}$ and $\mathbf{x}$ denote the noisy and clean images, respectively. (b) For a real-world point $\mathbf{r}$, a collection of raw images $\tilde{\mathcal{X}} = \{\tilde{\mathbf{x}}_1, \tilde{\mathbf{x}}_2\}$ records its intensity at pixel coordinates $\{\mathbf{p}_1, \mathbf{p}_2\}$, influenced by noise. The optimal target of 3DGS for this point, denoted as $\hat{\mathbf{x}}(\mathbf{p}) = \mathbb{E}_{\tilde{\mathbf{x}}(\mathbf{p}) \sim \tilde{\mathcal{X}}}$, has a discrepancy from the clean pixel intensity $\mathbf{x}(\mathbf{p})$. The variance of this discrepancy is detailed in Eq.~\eqref{eq:var_optimal_target}.
    }\label{fig:noise_model}
\end{figure}

\textbf{Noise model.}\label{sec:noise_model}
In the camera imaging process, depicted in Fig.~\ref{fig:noise_model_pmn}, the raw image $\tilde{\mathbf{x}}$ derived from the ideal clean raw image $\mathbf{x}$ inevitably contains noise $\mathbf{n}$. This relationship is expressed as:
\begin{equation}
    \tilde{\mathbf{x}} = \mathbf{x} + \mathbf{n}.
\end{equation}
Drawing on existing noise models~\cite{wei2020physics, feng2022learnability}, we decompose $\mathbf{n}$ into various noise components as:
\begin{equation}
    \mathbf{n} = \mathbf{n}_{shot} + \mathbf{n}_{read} + \mathbf{n}_{fp},
\end{equation}
where $\mathbf{n}_{shot}$, $\mathbf{n}_{read}$, and $\mathbf{n}_{fp}$ denote shot noise, read noise, and fixed pattern noise, respectively. A comprehensive description of each noise component is provided in the Supplemental Material. Given that $\mathbf{n}_{shot}$ can be approximated from the Poisson distribution $\mathcal{P}(\frac{\mathbf{x}}{k}) \cdot k - \mathbf{x}$ to a Gaussian distribution $\mathcal{N}(0, \mathbf{x} \cdot k)$, where $k$ represents the camera system gain, it can be seamlessly combined with $\mathbf{n}_{read}$, which also follows a Gaussian distribution $\mathcal{N}(0, \sigma^2_{read})$. This integration yields a heteroscedastic Gaussian noise model:
\begin{equation}
    \label{eq:noise_model_hg}
    \mathbf{n}_{hg} \sim \mathcal{N} (0, \boldsymbol{\sigma}^2_{hg}), \quad\boldsymbol{\sigma}^2_{hg} = \sigma^2_{read} + \mathbf{x} \cdot k.
\end{equation}
Thus, the overall noise model simplifies to:
\begin{equation}
    \label{eq:noise_model}
    \mathbf{n} = \mathbf{n}_{hg} + \mathbf{n}_{fp}.
\end{equation}
Utilizing Eq.~\eqref{eq:noise_model} as the prior for noise modeling allows for the potential separation of the clean image $\mathbf{x}$ from noise corruption, enabling more precise 3DGS reconstructions.

\textbf{3D Gaussian Splatting.}
3D Gaussian Splatting (3DGS)~\cite{kerbl3Dgaussians} models a 3D scene using a collection of 3D Gaussians, \textit{i.e.}, $\mathcal{G}=\{\mathbf{g}_1, \ldots, \mathbf{g}_{M}\}$, which are rendered from various viewpoints via a differentiable splatting and tile-based rasterization process. Each Gaussian $\mathbf{g}_i$ is initialized from Structure-from-Motion (SfM) and characterized by a 3D covariance matrix $\mathbf{\Sigma}_i = \mathbb{R}^{3\times3}$, position vectors $\boldsymbol{\mu}_i \in \mathbb{R}^3$, a color vector $\mathbf{c}_i \in \mathbb{R}^3$, and opacity $o_i$. To maintain the positive semi-definiteness of $\mathbf{\Sigma}_i$, it is parameterized as $\mathbf{\Sigma}_i = \mathbf{R}_i\mathbf{S}_i\mathbf{S}_i^T\mathbf{R}_i^T$, where $\mathbf{R}_i \in \mathbb{R}^{3\times3}$ and diagonal matrix $\mathbf{S}_i \in \mathbb{R}^{3\times3}$ represent rotation and scaling matrices, respectively. To render a pixel $\hat{\mathbf{x}} ({\mathbf{p}})$ at a pixel position $\mathbf{p}$ in the reconstructed image $\hat{\mathbf{x}}$ from a selected viewpoint within the camera projection $\mathcal{J}$, 3DGS utilizes a differentiable splatting operation to splat 3D Gaussians onto the 2D surface as follows:
\begin{equation}
\begin{aligned}
    \label{eq:3dgs}
    \mathbf{\hat{x}} (\mathbf{p}) &= \sum_{i=1}^{M} \mathbf{c}_i \alpha_i(\mathbf{p}) \prod_{j=1}^{i-1} (1-\alpha_j(\mathbf{p})), \\
    \alpha_i(\mathbf{p}) &= o_i \cdot \exp\left(-\frac{1}{2}(\mathbf{p} - \mathcal{J}(\boldsymbol{\mu}_i))^T\mathcal{J}(\mathbf{\Sigma}_i^{-1})(\mathbf{p} - \mathcal{J}(\boldsymbol{\mu}_i))\right).
\end{aligned}
\end{equation}
All attributes, \textit{i.e.}, $\mathbf{g}_i = \{\boldsymbol{\mu}_i, \mathbf{R}_i, \mathbf{S}_i, \mathbf{c}_i, o_i\}$, are learnable and optimized through the loss function:
\begin{equation}
    \mathcal{L}_{\text{3DGS}}(\hat{\mathbf{x}},\mathbf{x}) = (1-\lambda)\mathcal{L}_1(\hat{\mathbf{x}},\mathbf{x})+\lambda{L}_{\text{D-SSIM}}(\hat{\mathbf{x}},\mathbf{x}).
\end{equation}

\textbf{RawNeRF scaling loss.}
As discussed in RawNeRF~\cite{rawnerf}, the dynamic range of colors in a raw image is vast, which complicates the application of standard $\mathcal{L}_1$ or $\mathcal{L}_2$ loss. These losses tend to overemphasize errors in brighter areas, leading to tone-mapped images with poorly rendered, low-contrast dark regions. To address this, RawNeRF proposes a scaling loss as:
\begin{equation}
    \mathcal{L}_{\text{RawNeRF}}(\hat{\mathbf{x}},\mathbf{x}) = {\left( \frac{\hat{\mathbf{x}} - \mathbf{x}}{\text{sg}(\hat{\mathbf{x}}) + \epsilon} \right)}^2,
\end{equation}
where $\text{sg}(\cdot)$ is the stop gradient function, and $\epsilon$ is a small constant to prevent division by zero.

\section{Methodology}

\begin{figure}
    \centering
    \includegraphics[width=0.96\textwidth]{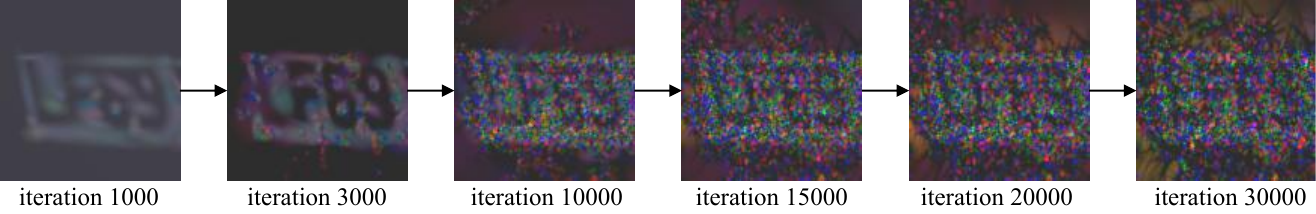}
    
    \caption{
        Visualization of the 3DGS test view changes across optimization iterations. Initially, the 3DGS model fits the clean signal (at 1,000 and 3,000 iterations). However, as the iterations progress (from 10,000 to 30,000), the model starts to overfit the noise.
    }\label{fig:dip}
    
    \vspace{-1em}
\end{figure}


\begin{figure}[t]
    \centering
    \includegraphics[width=0.96\textwidth]{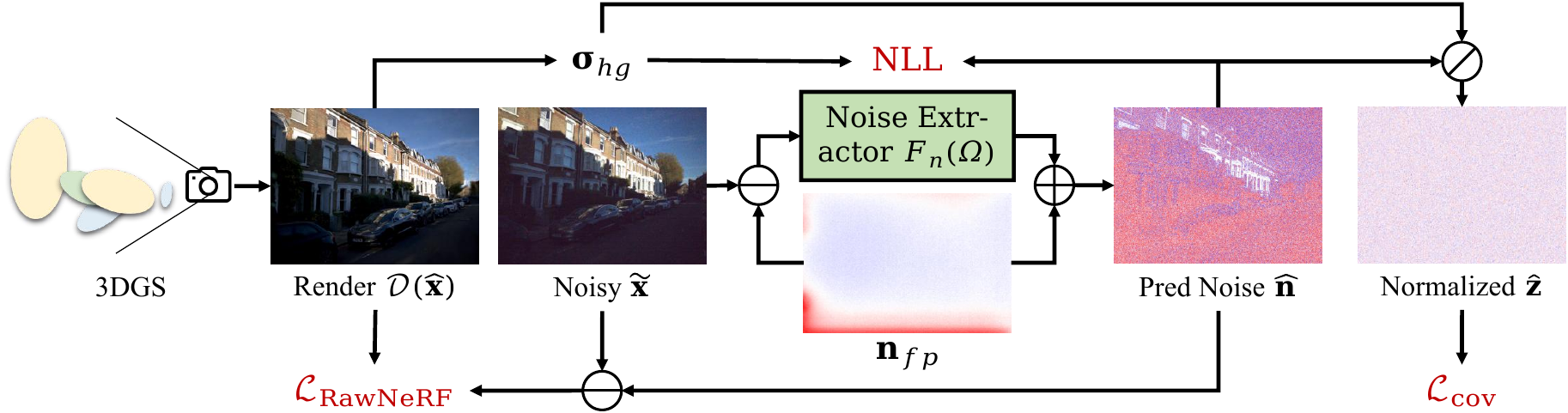}
    
    \caption{
        Illustration of the noise-robust reconstruction loss, $\mathcal{L}_{\text{nrr}}$, which comprises three components: the reconstruction loss $\mathcal{L}_{\text{RawNeRF}}$, the negative likelihood loss (NLL), and the covariance loss $\mathcal{L}_{\text{cov}}$. A noisy raw image, $\tilde{\mathbf{x}}$, is first input to the noise extractor $F_n(\cdot;\Omega)$ to estimate the noise, $\hat{\mathbf{n}}$. The estimated noise $\hat{\mathbf{n}}$ is then used to calculate the NLL loss relative to the noise distribution. After that, the normalized noise, $\hat{\mathbf{z}}$, undergoes a covariance loss, $\mathcal{L}_{\text{cov}}$, to minimize spatial dependencies among noise components. Finally, the reconstruction loss, $\mathcal{L}_{\text{RawNeRF}}$, is computed between the rendered distorted image $\mathcal{D}(\hat{\mathbf{x}})$ and the pseudo clean image $\tilde{\mathbf{x}}-\hat{\mathbf{n}}$.
    }\label{fig:framework}
    
    \vspace{-1em}
\end{figure}

\subsection{Motivation}

Consider a set of noisy raw images, denoted as $\tilde{\mathcal{X}} = \{\tilde{\mathbf{x}}_1, \ldots, \tilde{\mathbf{x}}_{N}\}$, which are captured from different viewpoints using the same camera settings. For any real-world point, \textit{e.g.}, $\mathbf{r}$, that is captured by all cameras, its intensity is recorded at a specific pixel coordinate $\mathbf{p}_i$ in each raw image $\tilde{\mathbf{x}}_i$, with the corresponding pixel value $\tilde{\mathbf{x}}_i({\mathbf{p}_i})$. As discussed in Sec.~\ref{sec:noise_model}, the noisy pixel value is the sum of the noise-free value and the noise at that location, \textit{i.e.}, $\tilde{\mathbf{x}}_i({\mathbf{p}_i}) = \mathbf{x}({\mathbf{p}}) + \mathbf{n}_i({\mathbf{p}_i})$. Given the same camera settings and real-world point $\mathbf{r}$, the clean value $\mathbf{x}({\mathbf{p}})$ remains constant across images.

To further explore the noise impact on 3D Gaussian Splatting (3DGS)~\cite{kerbl3Dgaussians} optimization, we simplify our analysis by utilizing an $\mathcal{L}_2$ loss function. The optimal target for the 3DGS output $\hat{\mathbf{x}}(\mathbf{p})$, which corresponds to different positions $\mathbf{p}_i$ in each image $\tilde{\mathbf{x}}_{i}$ after the splatting process with camera projection $\mathcal{J}_i$, is the following:
\begin{equation}
    \hat{\mathbf{x}}({\mathbf{p}}) = 
    \arg\min_{\hat{\mathbf{x}}({\mathbf{p}})} 
    \sum_{i=1}^{N} 
    \mathcal{L}_2(
        \hat{\mathbf{x}}({\mathbf{p}}), 
        \tilde{\mathbf{x}}_i({\mathbf{p}_i})) 
    = \mathbf{x}(\mathbf{p}) 
    + \frac{1}{N} \cdot \sum_{i=1}^{N} \mathbf{n}_i({\mathbf{p}_i}). 
\end{equation}

Considering a real-world flat surface around point $\mathbf{r}$ where intensities remain consistent, independent noise will cause variations in the optimal target across the surface as:
\begin{equation}\label{eq:var_optimal_target}
    \text{Var}(\hat{\mathbf{x}}({\mathbf{p}})) = \frac{1}{N^2} \cdot \text{Var}(\mathbf{n}_i({\mathbf{p}_i})) = \frac{1}{N^2} \cdot \boldsymbol{\sigma}_{hg}^2.
\end{equation} 

With the same camera settings, it is evident that the variance of the optimal target $\hat{\mathbf{x}}(\mathbf{p})$ is inversely proportional to the number of training views $N$. 
Since 3D Gaussians are splatted according to Eq.~\eqref{eq:3dgs}, the splatting process tends to produce numerous thin, flat Gaussian shapes to compensate for this variance. 
With fewer viewpoints, which is a common setting in real-world applications, these elongated Gaussians dominate, as illustrated in Fig.~\ref{fig:cover}. This dominance not only degrades the reconstruction quality but also affects rendering speed, as the performance is directly linked to the number of Gaussian points $M$.

Although 3DGS decomposes the variance on a flat surface into numerous elongated Gaussians, we observed that these Gaussian points tend to collapse and are subsequently divided during the later stages of the optimization process.
In this process, the low-frequency signals are the first to be approximated before noise, as depicted in Fig.~\ref{fig:dip}, which aligns with Deep Image Prior (DIP)~\cite{ulyanov2018deep}, in which the low-frequency image components are typically reconstructed before noise. Based on such an observation, we propose incorporating a noise model as a prior in Sec.~\ref{sec:noise_regularization} to relax the constraints in the 3DGS optimization framework.

\subsection{Lens distortion correction}\label{sec:lens_distortion}

Unlike NeRF~\cite{mildenhall2021nerf}, which employs ray-tracing directly from image data, 3DGS maps 3D Gaussians onto a 2D image plane through a splatting operation. To enhance processing efficiency, 3DGS adopts the straightforward pinhole camera model. However, this model does not account for the nonlinear distortions—primarily radial—that are typical with real-world camera lenses. These distortions are more significant in raw images, where the typical lens corrections provided by ISP are absent.

To effectively correct for both radial and tangential lens distortions, we employ a distortion mapping function $\mathcal{D}(\cdot)$. This function transforms the undistorted image coordinates $(x, y)$ into their corresponding distorted coordinates $(x_d, y_d)$, based on the distortion model used in COLMAP~\cite{schonberger2016structure}. The distortion map is derived through an iterative Newton-Raphson method. This computation is performed once prior to the training phase, thereby not affecting the training duration. Detailed explanations of the distortion mapping process are provided in the Supplemental Material.






    

\subsection{Noise robust reconstruction loss}\label{sec:noise_regularization}

To mitigate overfitting to noise in 3DGS, we have revised the reconstruction loss function, now termed the noise-robust reconstruction loss function $\mathcal{L}_{\text{nrr}}$, as depicted in Fig.~\ref{fig:framework}. This function is defined as follows:
\begin{equation}
    \mathcal{L}_{\text{nrr}} (\mathcal{D}(\hat{\mathbf{x}}), \tilde{\mathbf{x}}) = \mathcal{L}_{\text{RawNeRF}}(\mathcal{D}(\hat{\mathbf{x}}), \tilde{\mathbf{x}} - \hat{\mathbf{n}}) + \lambda_{\text{nd}} \cdot \mathcal{L}_{\text{nd}}(\hat{\mathbf{n}}, \mathbf{n}),
\end{equation}
where $\mathcal{D}(\cdot)$ denotes the lens distortion mapping as described in Sec.~\ref{sec:lens_distortion}. The function $\mathcal{L}_{\text{nd}}(\hat{\mathbf{n}}, \mathbf{n})$ measures the noise divergence between the estimated noise $\hat{\mathbf{n}} \sim q_{\hat{\mathbf{n}}}$ and the physical noise model $\mathbf{n} \sim p_{\mathbf{n}}$. The term $\lambda_{\text{nd}}$ is a Lagrangian relaxation parameter that balances fidelity and noise reduction. To estimate the noise, we introduce a noise extractor $F_n(\cdot; \Omega)$, a neural network parameterized by $\Omega$, which predicts the noise $\hat{\mathbf{n}}$ from the input raw image $\tilde{\mathbf{x}}$ as:
\begin{equation}
    \hat{\mathbf{n}} = F_n(\tilde{\mathbf{x}}-\mathbf{n}_{fp};\Omega) + \mathbf{n}_{fp}.
\end{equation}
Here, we subtract the input by the fixed pattern noise $\mathbf{n}_{fp}$ before feeding it into the noise extractor $F_{n}(\cdot; \Omega)$, due to the consistent nature of fixed pattern noise across different captures.

For the noise divergence loss $\mathcal{L}_{\text{nd}}(\hat{\mathbf{n}}, \mathbf{n})$, we first use the NLL Loss to measure the divergence between the estimated noise and the physical noise in corresponding pixel coordinates. The NLL loss is defined as:
\begin{equation}
    \text{NLL}(\hat{\mathbf{n}}, \mathbf{n}) 
    = \mathbb{E}_{q_{\hat{\mathbf{n}}}}[-\log p_{\mathbf{n}}(\hat{\mathbf{n}})] , 
    \quad p_{\mathbf{n}} = \mathcal{N}(0, \boldsymbol{\sigma}_{hg}^2) + \mathbf{n}_{fp},
\end{equation}
where $\boldsymbol{\sigma}_{hg}^2 = \sigma_{read}^2 + \hat{\mathbf{x}} \cdot k$ is calculated according to Eq.~\eqref{eq:noise_model_hg}.

The NLL loss models the distribution of predicted noise across each pixel, but it does not address the inter-pixel correlations, which ideally should be minimal. To tackle this, we then standardize the predicted noise $\hat{\mathbf{n}}$ to a standard Gaussian distribution $\mathcal{N}(0, 1)$ and introduce a covariance loss $\mathcal{L}_{\text{cov}}$ to minimize spatial dependencies:
\begin{equation}
    \mathcal{L}_{\text{cov}} (\hat{\mathbf{n}}, \mathbf{n})
    = \mathbb{E}_{q_{\hat{Z}|\tilde{X}}}[I - \hat{\mathbf{z}}^T \hat{\mathbf{z}}], 
    \quad \hat{\mathbf{z}} = (\hat{\mathbf{n}}-\mathbf{n}_{fp}) / \boldsymbol{\sigma}_{hg},
\end{equation}
where $I$ is the identity matrix, ensuring the decorrelation of noise across pixels.

Consequently, the comprehensive noise divergence loss $\mathcal{L}_{\text{nd}}(\hat{\mathbf{n}}, \mathbf{n})$ is formulated as:
\begin{equation}
    \mathcal{L}_{\text{nd}}(\hat{\mathbf{n}}, \mathbf{n}) 
    = \text{NLL}(\hat{\mathbf{n}}, \mathbf{n})
    + \lambda_{\text{cov}} / \lambda_{\text{nd}} \cdot \mathcal{L}_{\text{cov}} (\hat{\mathbf{n}}, \mathbf{n}).
\label{eq:nd_loss}
\end{equation}

Finally, the complete noise-robust reconstruction loss can be formulated as:
\begin{equation}
    \mathcal{L}_{\text{nrr}} (\mathcal{D}(\hat{\mathbf{x}}), \tilde{\mathbf{x}}) = \mathcal{L}_{\text{RawNeRF}}(\mathcal{D}(\hat{\mathbf{x}}), \tilde{\mathbf{x}} - \hat{\mathbf{n}}) + \lambda_{\text{nd}} \cdot \text{NLL}(\hat{\mathbf{n}}, \mathbf{n}) + \lambda_{\text{cov}} \cdot  \mathcal{L}_{\text{cov}} (\hat{\mathbf{n}}, \mathbf{n}).
\label{eq:nrr_loss}
\end{equation}

\section{Experiments}

In this section, we introduce the experimental setup, followed by comparing our methods with existing baselines. Finally, we conduct ablation studies to facilitate a thorough discussion of our approach.

\begin{figure}
    \centering
    
    \begin{subfigure}[b]{0.46\textwidth}
        \centering
        \includegraphics[width=\textwidth]{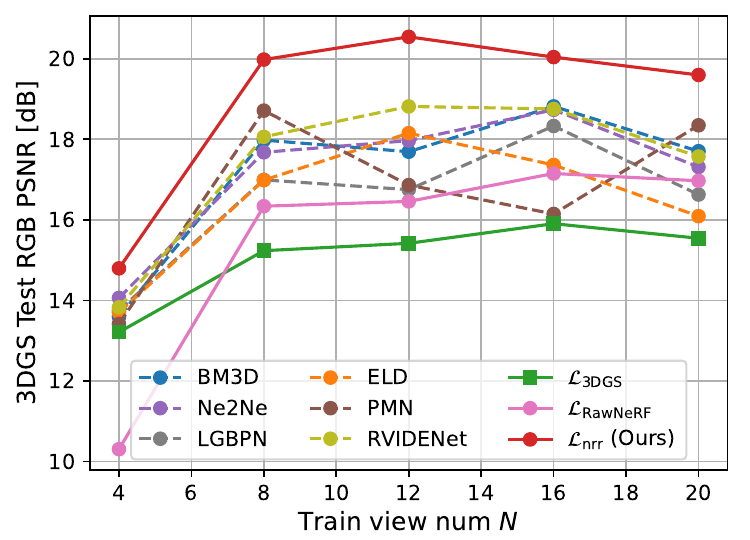}
        \caption{PSNR over the number of views in training.}
    \end{subfigure}\label{fig:few_shot_results_curve_psnr}
    \quad
    \begin{subfigure}[b]{0.46\textwidth}
        \centering
        \includegraphics[width=\textwidth]{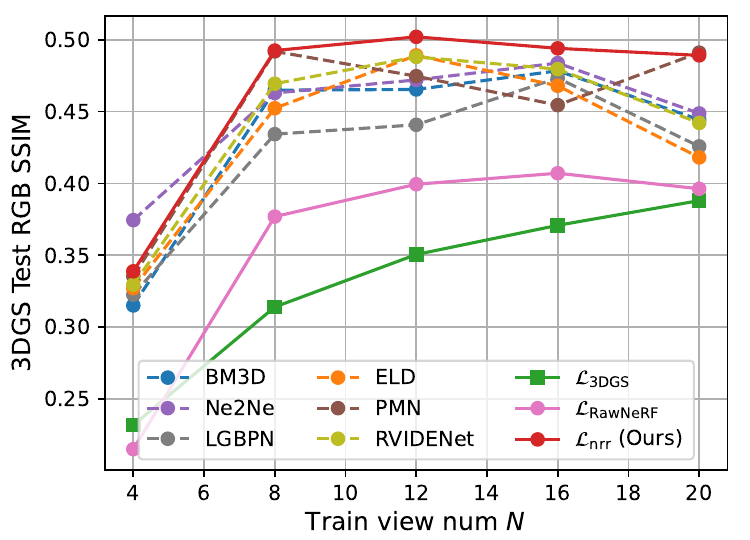}
        \caption{SSIM over the number of views in training.}
    \end{subfigure}\label{fig:few_shot_results_curve_ssim}
    
    \begin{subfigure}[b]{0.46\textwidth}
        \centering
        \includegraphics[width=\textwidth]{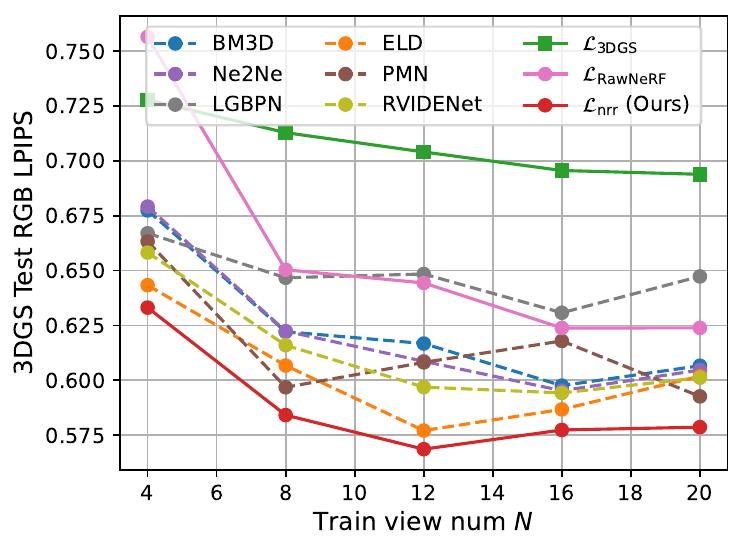}
        \caption{LPIPS over the number of views in training.}
    \end{subfigure}\label{fig:few_shot_results_curve_lpips}
    \quad
    \begin{subfigure}[b]{0.46\textwidth}
        \centering
        \includegraphics[width=\textwidth]{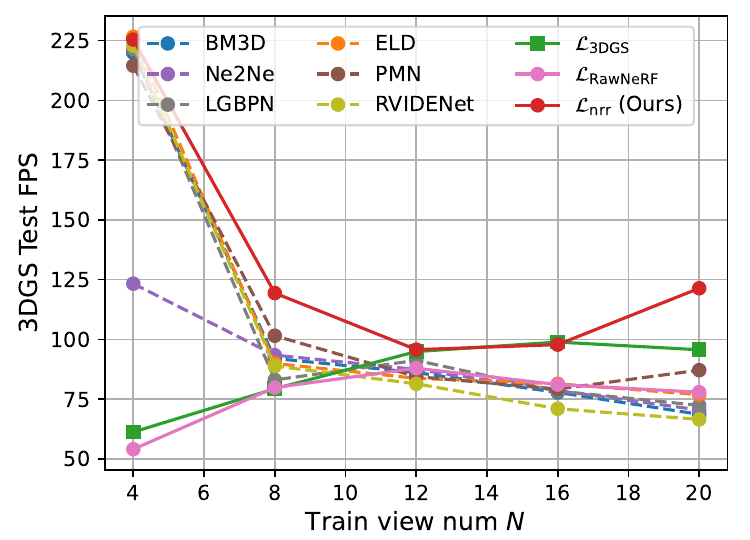}
        \caption{FPS over the number of views in training.}
    \end{subfigure}\label{fig:few_shot_results_curve_fps}

    \caption{
        Comparative evaluation of various baselines and our method on rendering quality and speed in limited views training settings. The two-stage denoiser + 3DGS methods are represented by dotted lines, while training on RGB images is indicated by square markers. All metrics are evaluated on test views within the RGB domain.
    }\label{fig:few_shot_results_curve}
    \vspace{-1em}
\end{figure}

\subsection{Implementation details}

Our framework is built upon Scaffold-GS~\cite{lu2023scaffold}. Considering the challenges associated with training $F_n(\cdot, \Omega)$ from scratch within only 30,000 iterations, we opted to pretrain it on the SID dataset~\cite{chen2018learning}, which features paired images of noise and noise-free scenes. Notably, the SID dataset is captured using a Sony DSLR camera, distinctly different from the iPhone X used in the RawNeRF dataset~\cite{rawnerf} for evaluation (e.g., bit depth differences: 14 vs 12), thereby mitigating the risk of data leakage.

Noise extractor $F_n(\cdot, \Omega)$ is optimized using the Adam optimizer with an initial learning rate of $1 \times 10^{-4}$, reduced to $1 \times 10^{-5}$ after 25,000 iterations. The hyperparameters $\lambda_{nd}$ and $\lambda_{\text{cov}}$, as defined in Eq.~\eqref{eq:nrr_loss} and Eq.~\eqref{eq:nd_loss}, are set to 5 and 20 for full-views settings, and 3 and 20 for limited views settings, respectively. All 3DGS models are trained on a single NVIDIA RTX 3090 GPU, following the default parameter settings of Scaffold-GS.

\subsection{Comparison with baselines}

\textbf{Baselines.} 
We first compare our method with LDR Scaffold-GS~\cite{lu2023scaffold}, which uses RGB images processed by an ISP as the optimal target, to demonstrate the benefits of training on raw images. The ISP settings are the same as those used in RawNeRF. We also benchmark our method against HDR Scaffold-GS optimized with the $\mathcal{L}_{\text{RawNeRF}}$ loss using raw images as a fair baseline. Additionally, we evaluate our approach against two-stage denoiser+3DGS pipelines optimized using the $\mathcal{L}_{\text{RawNeRF}}$ loss, where 3DGS is optimized on denoised raw images produced by the denoiser. These comparisons include traditional denoiser BM3D~\cite{dabov2007image}, state-of-the-art pretrained image denoisers such as ELD~\cite{wei2020physics} and PMN~\cite{feng2022learnability}, the pretrained video denoiser RViDeNet~\cite{yue2020supervised}, and advanced self-supervised methods like Neighbor2Neighbor~\cite{huang2021neighbor2neighbor} and LGBPN~\cite{wang2023lg} for thorough evaluations. For the pretrained supervised denoisers, we utilized the checkpoints provided by the authors. For the self-supervised methods, we initially pretrained the denoisers on the SID dataset~\cite{chen2018learning} and subsequently fine-tuned them on the training views. All compared methods apply the lens distortion function $\mathcal{D}(\cdot)$ to the rendered images to align with distorted raw images, ensuring a fair comparison.

\begin{figure}
    \centering
    \includegraphics[width=0.95\textwidth]{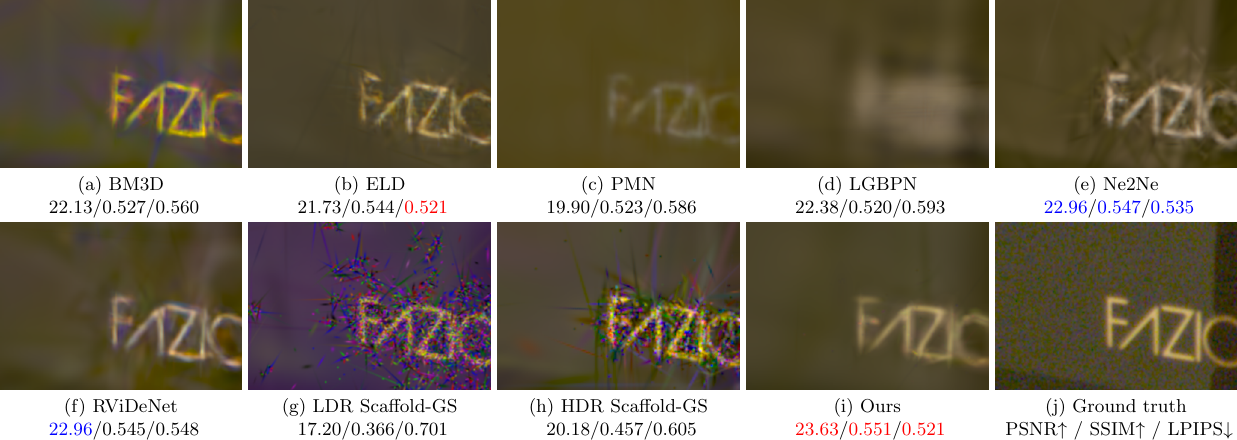}
    \caption{Visual comparison using ours and competing methods under the 12-view training settings.}\label{fig:visualization_fewshot_view}
    \vspace{-1em}
\end{figure}

\begin{figure}
    \centering
    \includegraphics[width=0.95\textwidth]{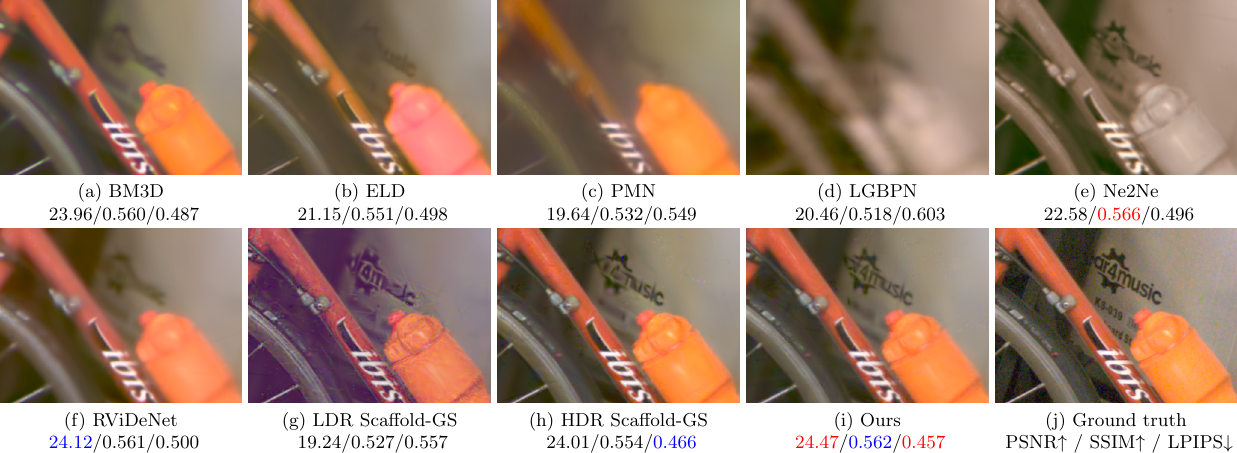}
    \caption{Visual comparison across various methods in full views (100-view) training settings.}\label{fig:visualization_full_view}
    \vspace{-1em}
\end{figure}

\textbf{Datasets and metrics.} 
We utilize the RawNeRF dataset~\footnote{\url{https://bmild.github.io/rawnerf/} \textcopyright\ google-research. Licensed under the Apache License, v2.0.\label{fn:license}}, which comprises raw images captured in dark scenes using an iPhone X with various ISO settings. For the full-views training setting, we adhere to the same train-test view splits as in RawNeRF. For limited views settings, we randomly select subsets (4, 8, 12, 16, 20 views) from the training views, while maintaining a consistent test view set across all experiments to ensure fair comparison. Following RawNeRF, we assess the rendering quality of various methods using PSNR, SSIM, and LPIPS metrics in the RGB domain. Additionally, we report the frames per second (FPS) to evaluate the real-time rendering capabilities of each method.

\textbf{Limited views training results.} In Fig.~\ref{fig:few_shot_results_curve}, we present the quantitative results of our approach compared to other baselines in limited views training settings. At first glance, it is evident that our method (indicated by red lines) significantly surpasses both the LDR Scaffold-GS, HDR Scaffold-GS, and two-stage methods across all view numbers. Compared to LDR Scaffold-GS (green lines), which utilizes RGB images processed by an ISP, our method achieves approximately a 4 dB improvement in PSNR, demonstrating the benefits of training with raw images. Additionally, when both are trained with raw images, our approach outperforms HDR Scaffold-GS (pink lines), optimized by $\mathcal{L}_{\text{RawNeRF}}$, by approximately 3 dB in PSNR across different training views, and about 4 times faster in FPS with limited 4-view settings, highlighting the efficiency of our noise robust reconstruction loss $\mathcal{L}_{\text{nrr}}$. Furthermore, our method also exceeds the two-stage denoiser + 3DGS pipelines by 1 dB, which may lose high-frequency details during the denoising process. Fig.~\ref{fig:visualization_fewshot_view} provides a visualization comparison. Notably, with 12 views, both LDR- and HDR- Scaffold-GS exhibit significant elongated Gaussian artifacts, while our method produces more accurate and detailed reconstructions.

\textbf{Full-views training results.} We also benchmarked our method against other baselines in full-views training settings. As shown in Table.~\ref{tab:full_view_results} and Fig.~\ref{fig:visualization_full_view}, our method consistently outperforms all other baselines across PSNR, SSIM, and LPIPS, even with a large number of training views. 
Besides, visualizations demonstrate that our method recovers images with better detail and color accuracy.

\begin{table}
    \small
    \centering
    \caption{Quantitative results of our approach and other baselines in full views training settings.}\label{tab:full_view_results}
 \scalebox{0.9}{
    \begin{tabular}{llccccc}
    \toprule
        \multirow{2}{*}{Method} & \multirow{2}{*}{Loss} & \multirow{2}{*}{Raw PSNR$\uparrow$} & \multicolumn{3}{c}{Affine-aligned RGB} & \multirow{2}{*}{FPS$\uparrow$} \\
        \cmidrule{4-6}
        & & & PSNR$\uparrow$ & SSIM$\uparrow$ & LPIPS$\downarrow$ \\
    \midrule
        BM3D~\cite{dabov2007image} & 
            \multirow{6}{*}{$\mathcal{L}_{\text{RawNeRF}}$}  &     
            \cellcolor{orange!40} 58.46 &       
            \cellcolor{orange!40} 22.70 &  
            \cellcolor{yellow!40} 0.526 &       
            0.521 & 
            76 \\
        ELD~\cite{wei2020physics} &  
            &     
            54.70 &      
            19.82 &       
            0.511 &     
            0.544 & 
            \cellcolor{yellow!40} 80 \\
        PMN~\cite{feng2022learnability} &  
            &     
            53.69 &      
            19.00 &       
            0.498 &     
            0.584 & 
            \cellcolor{red!40} 94 \\
        Neighbor2Neighbor~\cite{huang2021neighbor2neighbor} 
            &  
            &    
            \cellcolor{yellow!40} 58.40 &      
            22.22 &       
            \cellcolor{orange!40} 0.531 &     
            \cellcolor{yellow!40} 0.518 & 
            78 \\
        LGBPN~\cite{wang2023lg} &  
            &     
            53.62 &      
            19.09 &       
            0.481 &     
            0.624 & 
            \cellcolor{orange!40} 84 \\
        RViDeNet~\cite{yue2020supervised} &  
            &     
            57.59 &      
            22.05 &       
            0.518 &     
            0.548 & 
            74 \\
        \midrule
        LDR Scaffold-GS~\cite{lu2023scaffold} & 
            $\mathcal{L}_{\text{3DGS}}$ &     
            - &      
            17.34 &       
            0.486 &     
            0.622 & 
            56 \\ 
        HDR Scaffold-GS~\cite{lu2023scaffold} & 
            $\mathcal{L}_{\text{RawNeRF}}$ &
            58.08 &      
            \cellcolor{yellow!40} 22.69 &       
            0.521 &     
            \cellcolor{orange!40} 0.513 & 
            73 \\
        Ours & $\mathcal{L}_{\text{nrr}}$ &    
            \cellcolor{red!40} 59.49 &      
            \cellcolor{red!40} 23.53 &       
            \cellcolor{red!40} 0.535 &     
            \cellcolor{red!40} 0.499 & 
            \cellcolor{yellow!40} 80 \\
    \bottomrule
\end{tabular}}
\vspace{-1.3em}
\end{table}


\begin{table}
\centering

\begin{minipage}{0.5\textwidth}
    \small
    \setlength{\tabcolsep}{2.3pt}
    \caption{Ablation study on the lens distortion $\mathcal{D}(\cdot)$. All models are trained using full views.}\label{tab:ablation_distortion}
 \scalebox{0.9}{
    \begin{tabular}{ccccc}
    \toprule
        $\mathcal{D}(\cdot)$  & 
        Raw PSNR &
        RGB PSNR & 
        RGB SSIM & 
        RGB LPIPS \\
    \midrule
        w/o &
        59.33 &
        23.42 &
        0.531 &
        0.509
        \\
        w/ &
        \textbf{59.49} &
        \textbf{23.53} &
        \textbf{0.535} &
        \textbf{0.499} \\

    \bottomrule
    \end{tabular}}
    \normalsize
\end{minipage}
\hfill\hfill
\begin{minipage}{0.45\textwidth}
    \footnotesize
    \setlength{\tabcolsep}{5pt}
    \centering
    \caption{Ablation study on the training time.}\label{tab:ablatio_training_times}
 \scalebox{0.9}{
    \begin{tabular}{lcc}
    \toprule
        \multirow{2}{*}{Method}  & 
        Ne2Ne~\cite{huang2021neighbor2neighbor} & 
        LGBPN~\cite{wang2023lg} \\ 
        & 
        Scaffold-GS~\cite{lu2023scaffold} & 
        Ours \\ \midrule
        \multirow{2}{*}{Time cost} &
        2.5h &
        5.3h\\ 
        &
        1.6h&
        3.1h\\
    \bottomrule
    \end{tabular}}
    \normalsize
\end{minipage}
\vspace{-2.0em}
\end{table}

\begin{table}[H]
    \small
    \centering
    \caption{Ablation study on the selection of $\lambda_{\text{nd}}$ and $\lambda_{\text{cov}}$. All models are trained using full views.}\label{tab:ablatio_lambda}
 \scalebox{0.9}{
    \begin{tabular}{cccccc}
    \toprule
        $\lambda_{\text{nd}}$ & 
        $\lambda_{\text{cov}}$ & 
        Raw PSNR &
        RGB PSNR & 
        RGB SSIM & 
        RGB LPIPS \\
    \midrule
        0.5 &
        20 &
        56.36 &
        21.95 &
        0.508 &
        0.591 \\
        10 &
        20 &
        56.84 &
        22.18&
        0.509&
        0.539\\
        5 &
        0 &
        55.38 &
        22.41 &
        0.515 &
        0.523 \\
        5 &
        2 &
        57.10 &
        22.52 &
        0.516 &
        0.520 \\
        0.5 &
        50 &
        56.72 &
        21.98 &
        0.514 &
        0.576
        \\
        \textbf{5} &
        \textbf{20} &
        \textbf{59.49} &
        \textbf{23.53} &
        \textbf{0.535} &
        \textbf{0.499} \\

    \bottomrule
    \end{tabular}}
    \normalsize
    \vspace{-1.3em}
\end{table}

\subsection{Ablation study and discussions}
\textbf{Impact of lens distortion function $\mathcal{D}(\cdot)$.} As detailed in Table~\ref{tab:ablation_distortion}, the absence of lens distortion correction results in a marginal decrease in reconstruction quality. In RawNeRF datasets, this effect is negligible due to the minimal lens distortion inherent in the iPhone X. However, for cameras with more pronounced distortion characteristics, the impact could be significantly more substantial.

\textbf{Impact of hyperparameters.}
The hyperparameters $\lambda_{\text{nd}}$ and $\lambda_{\text{cov}}$ within the $\mathcal{L}_{\text{nrr}}$ play a pivotal role in striking the right balance between denoising efficacy and gradient accumulation. As delineated in Table~\ref{tab:ablatio_lambda}, an undersized $\lambda_{\text{nd}}$ leads to inadequate constraints on the noise model, culminating in imprecise noise extraction. On the flip side, an over-amplified $\lambda_{\text{nd}}$ engenders excessively strong gradients emanating from the noise regularization loss, which can detrimentally impact the optimization process of 3DGS. The parameter $\lambda_{\text{cov}}$, when assigned values that are either too low or too high, can incite similar challenges.

\textbf{Limitations.}\label{sec:limitations}
Our method has several limitations. The noise extractor and covariance loss computation require more training time compared to standard Scaffold-GS, as shown in Table~\ref{tab:ablatio_training_times}. Additionally, some high-frequency signal distributions are similar to noise, making it difficult to differentiate them using the distribution divergence loss. This can result in over-smoothing in full-views training settings, as discussed in the Supplemental Material.

\section{Conclusion}
In this work, we identify the unavoidable noise in raw images as a significant detriment to the rendering quality and speed of HDR 3DGS, particularly in scenarios with limited training views. We provide a comprehensive analysis of how noise affects the optimization of 3DGS, modeling its relationship with both the number of training views and the noise distribution. To tackle these challenges, we propose a novel self-supervised framework that incorporates a noise-robust reconstruction loss. This framework utilizes a physically-based noise model to simultaneously denoise and enhance the HDR representation within noisy raw images. Our approach markedly outperforms LDR/HDR 3DGS that employs 3DGS/RawNeRF loss, as well as both self-supervised and supervised pre-trained two-stage methods, in terms of rendering quality and speed across various training view counts.

\newpage
{
\small
\bibliographystyle{abbrv}
\bibliography{neurips_2024}
}


\newpage
\appendix

\section*{From Chaos to Clarity: 3DGS in the Dark — Supplemental Material}

In this supplement, we begin by discussing the broader impacts of our method. Next, we outline the distribution and calibration of the camera noise model. Following that, we elaborate on the lens distortion function $\mathcal{D}(\cdot)$. Finally, we address the failure cases of our method as mentioned in the limitations section.

\section{Broader Impacts}

The broader social impacts of our work can be summarized as follows: 
\begin{enumerate}
    \item \textbf{Enhanced Applications:} By significantly improving 3DGS performance in low-light conditions, our work has the potential to broaden the scope of its applications across various fields such as medical imaging, autonomous driving, and surveillance systems. These advancements could lead to better outcomes in scenarios where lighting is a critical factor.
    \item \textbf{Open Access and Collaboration:} In the spirit of fostering innovation and collaboration within the research community, we will release our code as open-source. This will allow other researchers and developers to build upon our work, potentially leading to further advancements and new applications.
\end{enumerate}

We have carefully considered the potential impacts of our work and do not foresee any serious negative consequences. Our contributions are intended to promote positive developments and collaboration in the field of 3DGS technology.

\section{Distribution of the camera noise model}

Building on the work of ELD~\cite{wei2020physics} and PMN~\cite{feng2022learnability}, the noise components in Eq.\eqref{eq:noise_model} adhere to specific distributions:
\begin{equation}
\begin{aligned}
\mathbf{n}_{shot} & \sim \mathcal{P}\left( \frac{\mathbf{x}}{k} \right) \cdot k - \mathbf{x},\\
\mathbf{n}_{read} & \sim \mathcal{N} ( 0, \sigma_{read}^2 ), \\
\mathbf{n}_{fp} & = \text{ISO} \cdot \mathbf{n}_{fp_k} + \mathbf{n}_{fp_b}, \label{eq:noise_pmn}
\end{aligned}
\end{equation}
where $k$ is the overall system gain associated with the ISO setting, and $\mathcal{P}$ and $\mathcal{N}$ represent Poisson and Gaussian distributions, respectively. The terms $n_{fp_k}$ and $n_{fp_b} \in \mathbb{R}^{H \times W}$ are pixel-wise dark frame noise components. Consistent with ELD\cite{wei2020physics} and PMN~\cite{feng2022learnability}, the relationships among $k$, $\sigma_{read}$, and ISO are expressed as:
\begin{equation}
\begin{aligned}
k &= a_{k} \cdot \text{ISO} + b_{k}, \\
\log\left(\sigma_{read}\right) &= a_{read} \cdot \log( k ) + b_{read},
\end{aligned}
\end{equation}
Parameters $n_{fp_k}$, $n_{fp_b}$, $a_{k}$, $b_{k}$, $a_{read}$, and $b_{read}$ are calibrated for each camera using a series of flat-frame and dark-frame images captured at various ISO levels.

\section{Calibration of the camera noise model parameters}

The calibration process comprises three steps:

\textbf{1. Calibrating $k_i$ for each ISO level:}
\begin{itemize}
    \item Capture 25 flat frames under consistent lighting for each exposure time $\text{Exp}_j$.
    \item Calculate the mean and variance for each color block, yielding 24 mean-variance pairs per exposure time.
    \item  With three exposure times per ISO, gather 72 mean-variance pairs per ISO.
    \item Model $\mathbf{n}_{shot}$ as $\mathcal{N}(\mathbf{x}, \mathbf{x} \cdot k)$, where $\mathbf{x}$ is the mean and $\mathbf{x} \cdot k$ the variance, and calibrate $k$ from the mean-variance relationship. Exclude points with mean values beyond 1/4 saturation due to clipping effects.
\end{itemize}

\textbf{2. Calibrating $\mathbf{n}_{{fp}_i}$ and $\sigma_{{read}_i}$:}

\begin{itemize}
    \item Capture 100 dark frames at each ISO in a dark room.
    \item The mean of these dark frames gives $\mathbf{n}_{fp_i}$, representing fixed pattern noise.
    \item Subtract $\mathbf{n}_{fp_i}$ from all dark frames and calculate variance across the total frame for $\mathbf{n}_{read}$.
\end{itemize}

\textbf{3. Fitting ISO-related parameters:}

\begin{itemize}
    \item Repeat the above steps for different ISO levels to obtain a set of parameters $\mathbf{n}_{{fp}_i}$, $\sigma_{{read}_i}$.
    \item Fit $a_k$, $b_k$, $n_{{fp}k}$, $n{fp_b}$, $a_{read}$, $b_{read}$ based on these parameters and equations Eq.\ref{eq:noise_pmn}.
\end{itemize}

\section{Lens distortion mapping process}

To align the distorted coordinates with their accurate positions, we compute a reverse mapping from $(x_d, y_d)$ to $(x, y)$ using a Newton-Raphson iterative method. This method minimizes the residuals between the distorted and true coordinates until the adjustments are beneath a predefined accuracy threshold. Crucially, this distortion mapping is computed once before the training process, thereby minimizing its impact on computational efficiency.

using the following equations:

\begin{equation}
    \begin{aligned}
        x_d &= x \left(1 + k_1 r^2 + k_2 r^4 + k_3 r^6 + k_4 r^8\right) + 2p_1 xy + p_2\left(r^2 + 2x^2\right), \\
        y_d &= y \left(1 + k_1 r^2 + k_2 r^4 + k_3 r^6 + k_4 r^8\right) + p_1\left(r^2 + 2y^2\right) + 2p_2 xy,
    \end{aligned}
\end{equation}
    
where $r = \sqrt{x^2 + y^2}$ represents the radial distance from the center, and $k_1, k_2, k_3, k_4$, $p_1, p_2$ are the coefficients for radial and tangential distortions, respectively.

Since $x$ and $y$ are not integers, we use bilinear interpolation to obtain the values of $x_d$ and $y_d$.

\section{Failure case}
\begin{figure}[t]
    \centering
    
    \begin{subfigure}[b]{0.48\textwidth}
        \centering
        \includegraphics[width=\textwidth]{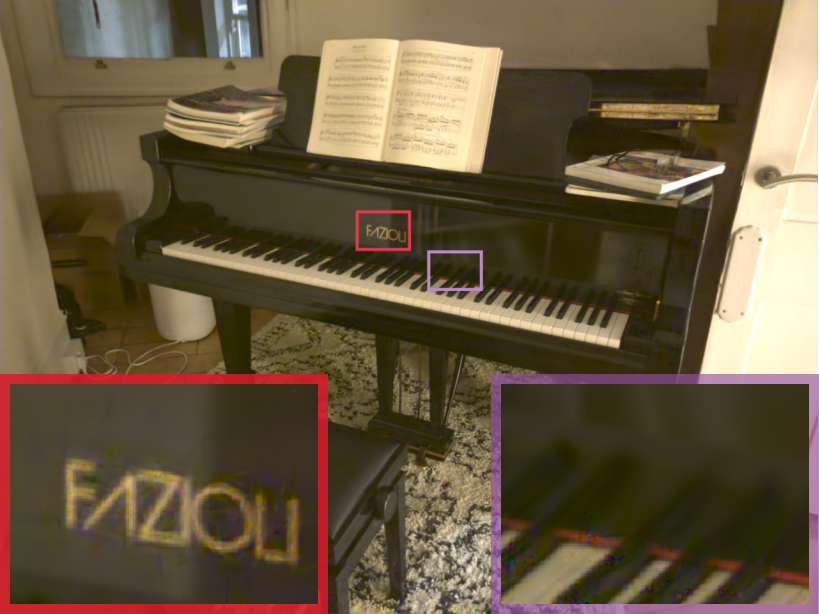}
        \caption{HDR Scaffold-GS (PSNR: 24.59 dB)}\label{fig:supplementary_rawnerf}
    \end{subfigure}
    \hfill
    \begin{subfigure}[b]{0.48\textwidth}
        \centering
        \includegraphics[width=\textwidth]{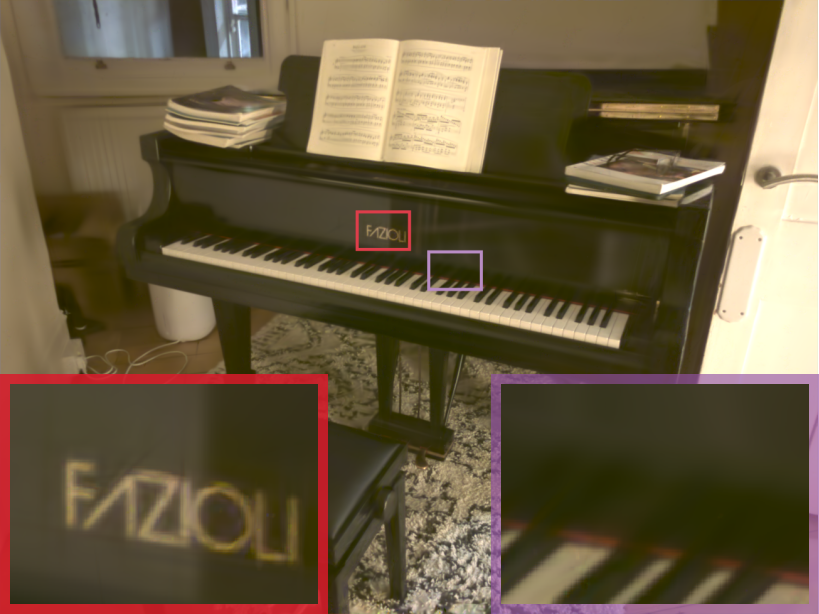}
        \caption{Ours (PSNR: 24.85 dB)}\label{fig:supplementary_ours}
    \end{subfigure}

    \caption{
        Visualization of our failure cases on the RawNeRF dataset~\cite{rawnerf}. Despite achieving a higher PSNR than HDR Scaffold-GS by eliminating noise-induced artifacts, our approach resulted in overly smooth outcomes in areas with reflections.
    }\label{fig:supplementary}
\end{figure}

Upon examination, we have observed that certain high-frequency signals exhibit a distribution pattern that closely resembles noise, making it difficult for the noise divergence loss to differentiate between them effectively. Consequently, this has resulted in overly smooth outputs in some areas rich in detail, particularly when training with full views.

Differentiating high-frequency signals from noise is an intricate challenge. We acknowledge the complexity of this issue and are committed to further exploring potential solutions in our future research endeavors.

\end{document}